\documentclass[final,5p,times,twocolumn]{elsarticle}
\usepackage{epsfig}
\usepackage{amssymb}
\usepackage{amsthm}
\usepackage{graphicx}
\usepackage{natbib}
\usepackage{xcolor}
\usepackage{multirow}
\usepackage{hyperref}
\usepackage{amsmath}

\makeatletter
\def\ps@pprintTitle{%
  \let\@oddhead\@empty
  \let\@evenhead\@empty
  \def\@oddfoot{\reset@font\hfil\thepage\hfil}
  \let\@evenfoot\@oddfoot
}
\makeatother

\begin{document}

\begin{frontmatter}

\title{ {\bf Prediction of activity coefficients in water-methanol mixtures using a generalized Debye-H{\"u}ckel model } }

\author[1]{Chin-Lung Li}
\ead{chinlungli@mx.nthu.edu.tw}

\author[1]{Shu-Yi Chou}
\ead{sam86928@gmail.com}

\author[1]{Jinn-Liang Liu\corref{cor1}}
\address[1]{Institute of Computational and Modeling Science, National Tsing Hua University, Hsinchu 30013, Taiwan}
\ead{jlliu@mx.nthu.edu.tw}
\cortext[cor1]{Corresponding author.}

\begin{abstract}
We propose a generalized Debye-H{\"u}ckel model from Poisson-Fermi theory to predict the mean activity coefficient of electrolytes in water-methanol mixtures with arbitrary percentage of methanol from 0 to 100\%. The model applies to any number of ionic species and accounts for both short and long ion-ion, ion-water, ion-methanol, and water-methanol interactions, the size effect of all particles, and the dielectric effect of  mixed-solvent solutions. We also present a numerical algorithm with mathematical and physical details for using the model to fit or predict experimental data. The model has only 3 empirical parameters to fit the experimental data of NaF, NaCl, and NaBr, for example, in pure-water solutions. It then uses another 3 parameters to predict the activities of these salts in mixed-solvent solutions for any percentage of methanol. Values of these parameters show mathematical or physical meaning of ionic activities under variable mixing condition and salt concentration. The algorithm can automatically determine optimal values for the 3 fitting parameters without any manual adjustments.
\end{abstract}

\end{frontmatter}

%
%
\section{Introduction}
Water and alcohol are ubiquitous and complicated liquids \cite{Stu12,Fra96,Kir07}. With salts, they pose many challenges to thermodynamic modeling for a great variety of applications in a wide range of systems and conditions \cite{Voi11,Row15,Ver16,Wil17,Kon18,Kon20}. One of the major difficulties for numerous models \cite{Kon20} is to deal with the combinatorial explosion of empirical parameters up to tens of thousands \cite{Voi11} to calculate activity coefficients of electrolyte solutions with different compositions at variable temperature and pressure. Even worse, many parameters do not have physical meaning \cite{Fra10} or offer mathematical hint \cite{Voi11,Li20} to use.

Generalized Debye-H{\"u}ckel (DH) models \cite{Li20,Liu15,Liu18,Liu19} from a Poisson-Fermi (PF) theory \cite{Liu13,Liu13a,Liu14,Liu20} developed recently can ease some of these difficulties. These models use only 3 empirical parameters having both physical and mathematical properties to well fit
experimental activity data of multi-component electrolyte solutions in a range of concentrations, temperatures, and pressures. It is shown in \cite{Li20} that the generalized DH model differs much from H\"uckel's model \cite{Huc25} (and numerous DH models extended from it since 1925) as their approximations of Born solvation energies are inverse of each other in terms of parameters, which explains why extended DH models need more parameters generally without physical meaning. The PF theory treats ions and water (solvent) molecules of any volume and shape with interstitial voids, and accounts for polarization of water, both short and long ranges of ion-ion and ion-water interactions and correlations, and the non-uniform dielectric response (permittivity) of electrolyte solutions.

We propose here a generalized DH model to predict mean activity coefficients of electrolytes in water-methanol mixtures with any percentage (mole fraction) of methanol $x$ in [0, 1]. The model first uses 3 parameters $\alpha_j^{\text{H$_2$O}}$ for $j=1, 2, 3$ to best fit the experimental activity data of NaF \cite{Her03}, NaCl \cite{Bas96}, and NaBr \cite{Han93}, for example, in pure-water solutions (i.e., $x=0$) using the method of least squares. It then uses another 3 parameters $\Delta \alpha_j$ to predict the activities in mixed solutions for any arbitrary $x \ne 0$. The parameters $\alpha_j^{\text{H$_2$O}}$ define a factor function $\theta(I)$ of the variable ionic strength $I$ of the solution that in turn modifies the experimental Born radius $R_i^0$ \cite{Faw04,Pli13,Val15} of an ion $i$ in pure solvent (i.e., $I=0$) to an unknown Born radius $R_i^{Born} = \theta(I) R_i^0$ for any $I \ne 0$. The other $\Delta \alpha_j$ are defined by $\alpha_j^x = (1-x)\alpha_j^{\text{H$_2$O}} + x \alpha_j^\text{MeOH} = \alpha_j^{\text{H$_2$O}} + x \Delta \alpha_j$ for any mixing $x$. Therefore, all these 6 parameters have clear physical meaning in terms of Born energy. We also provide numerical evidence that their values offer novel hints for future studies on different solutions for which a numerical algorithm is given to show how to implement the model with details. This predictive model and the algorithm can be straightforwardly applied to electrolyte solutions with multi-valent ions, mixed salts, variable temperature, and variable pressure \cite{Li20}.

%
%
\section{Theory and algorithm}

For $K$ species of ions in water-methanol mixed solvents, the entropy model
proposed in \cite{Liu14,Liu20} treats ions, water (denoted by $K+1$), and methanol ($K+2$) as nonuniform spheres with interstitial voids ($K+3$). The total volume $V$ of the system is%
\begin{equation}
V=\sum_{i=1}^{K+2}v_{i}N_{i}+V_{K+3}, \label{2.1}%
\end{equation}
where $v_{i}$ is the volume of each
$i^{\text{th}}$ species particle, $N_{i}$ is the total number of all
$i^{\text{th}}$ species particles, and $V_{K+3}$ denotes the total volume of
all the voids. In bulk solutions, we have the bulk concentrations $C_{i}%
^{B}=\frac{N_{i}}{V}$ and the bulk volume fraction of voids $\Gamma^{B}%
=\frac{V_{K+3}}{V}$. Dividing (\ref{2.1}) by $V$, we get $\Gamma^{B}%
=1-\sum_{i=1}^{K+2}v_{i}C_{i}^{B}$. If the system is spatially inhomogeneous
with variable electric or steric fields, as in realistic systems, the constant
$C_{i}^{B}$ then changes to a function $C_{i}(\mathbf{r})$ so does $\Gamma
^{B}$ to a void volume function $\Gamma(\mathbf{r)}=1-\sum_{i=1}^{K+2}%
v_{i}C_{i}(\mathbf{r})$ in the solvent domain $\Omega_{s}$.

It is shown in \cite{Liu20} that the distribution (concentration) of
particles%
\begin{equation}
C_{i}(\mathbf{r})=C_{i}^{B}\exp\left(  -\beta_{i}\phi(\mathbf{r})+\frac{v_{i}%
}{v_{0}}S(\mathbf{r})\right)  \text{, \ \ }S(\mathbf{r}%
)=\ln\left(  \frac{\Gamma(\mathbf{r)}}{\Gamma^{B}}\right), \label{2.2}%
\end{equation}
is of Fermi-like type, i.e., $C_{i}(\mathbf{r})<\frac{1}{v_{i}}$ for any arbitrary (or even infinite)
electric potential $\phi(\mathbf{r})$ at any $\mathbf{r\in}$ $\Omega_{s}$ for
all $i=1,$ $\cdots,$ $K+2$, where $\beta_{i}=q_{i}/k_{B}T$ with $q_{i}$ being
the charge on species $i$ particles and $q_{K+1}=q_{K+2}=0$, $k_{B}$ is the Boltzmann
constant, $T$ is an absolute temperature, and $v_{0}$ is a unit volume. The steric potential $S(\mathbf{r})$ is an entropic measure of crowding or emptiness of
particles at $\mathbf{r}$
\cite{Liu13,Liu14,Liu20}. If $\phi(\mathbf{r})=0$, then $\Gamma
(\mathbf{r)}=\Gamma^{B}$ and hence $S(\mathbf{r})=0$. The factor
$v_{i}/v_{0}$ in (\ref{2.2}) shows that the steric energy $\frac{-v_{i}}%
{v_{0}}S(\mathbf{r})k_{B}T$ of a type $i$ particle at $\mathbf{r}$
depends not only on the steric potential $S(\mathbf{r})$ but also on its
volume $v_{i}$ similar to the electric energy $\beta_{i}\phi(\mathbf{r}%
)k_{B}T$ depending on both $\phi(\mathbf{r})$ and $q_{i}$
\cite{Liu20}.

The activity coefficient $\gamma_i$ of an ion of species $i$ in electrolyte solutions describes the deviation of chemical potential of the ion from ideality ($\gamma_i = 1$). The excess chemical potential $\mu_i^{ex} = k_B T \ln \gamma_i$ can be calculated by \cite{Liu15}
\begin{equation}
\mu_i^{ex} = \dfrac{1}{2}q_i\phi(\textbf{0})-\dfrac{1}{2}q_i\phi^0(\textbf{0}), \label{formula:excess chemical potential}
\end{equation}
 where $\phi(\textbf{r})$ is the electric potential generated by the ion $i$ in the system domain $\overline{\Omega} = \overline{\Omega}_i\cup\overline{\Omega}_{sh}\cup\overline{\Omega}_s$ shown in Fig. 1, $\overline{\Omega}_i$ is the spherical domain occupied by the ion, $\overline{\Omega}_{sh}$ is the solvation shell domain of the ion, $\overline{\Omega}_s$ is the rest of solvent domain, $\textbf{0}$ denotes the center of the ion, and $\phi^0(\textbf{r})$ is a potential function when the solvent domain is ideal, i.e., $C_{j}^{B}=0$ for all $j$.

\begin{figure}[h]
\begin{center}
\includegraphics[width=3.3in]{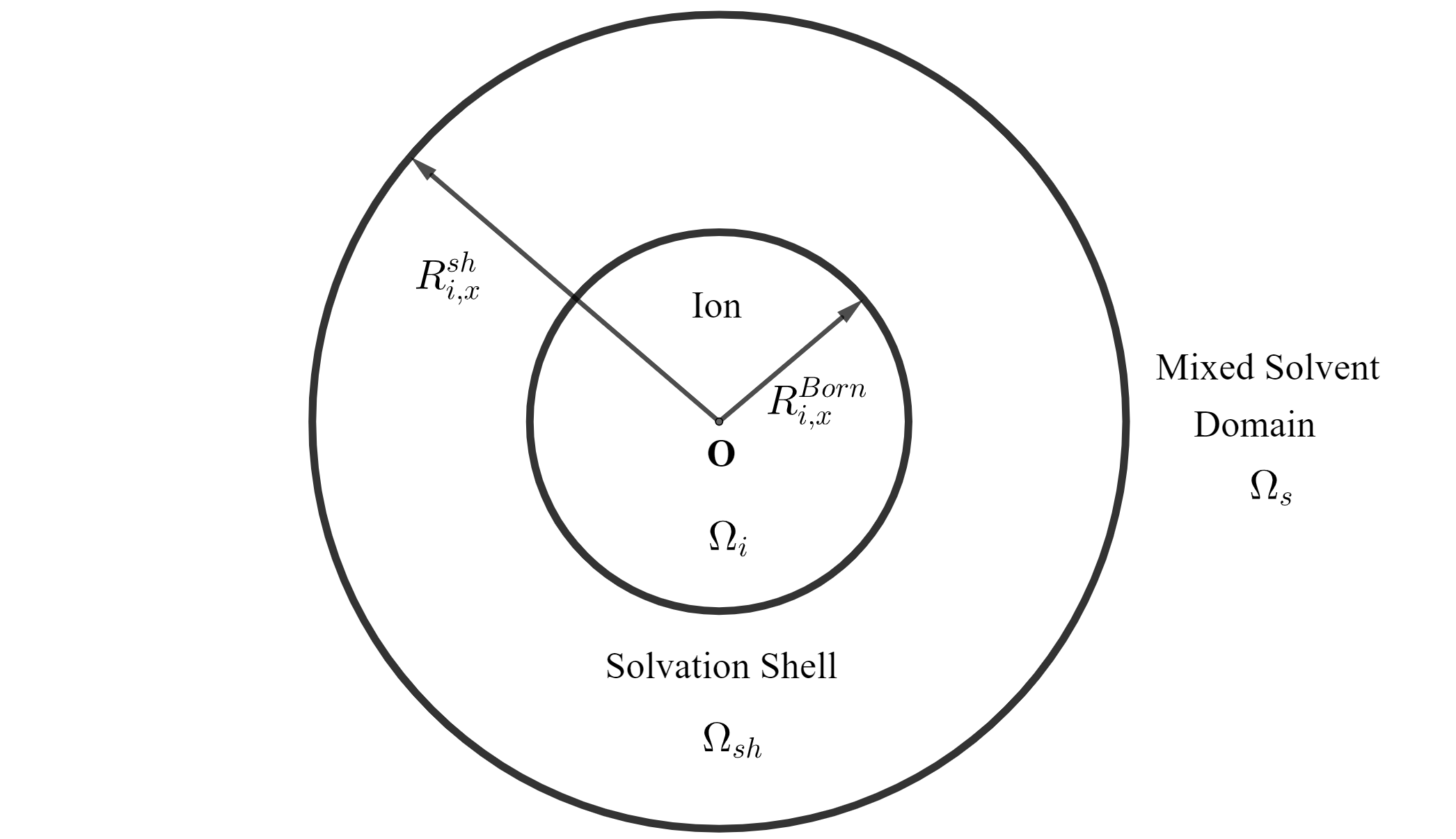}
\caption{The model domain $\Omega$ is partitioned into ion domain $\Omega_i$ (with radius $R_{i,x}^{Born}$), the solvation shell domain $\Omega_{sh}$ (with radius $R_{i,x}^{sh}$), and the remaining mixed solvent domain $\Omega_s$.}
\end{center}
\end{figure}

The potential function $\phi(\textbf{r})$ can be found by solving the PF equation
\cite{Li20,Liu15,Liu18,Liu19}
\begin{equation}
\epsilon_s(l_c^2 \nabla^2-1) \nabla^2 \phi(\textbf{r}) =\sum_{i = 1}^{K}q_iC_i(\textbf{r}) \text{ in } \Omega_s \label{formula:PF equation}
\end{equation}
and the Laplace equation
\begin{equation}
-\nabla^2 \phi(\textbf{r}) = 0 \text{ in }  \Omega_i \cup \Omega_{sh}, \label{formula:Laplace equation}
\end{equation}
where $\epsilon_s = \epsilon_0\epsilon_x$ is the permittivity of a {\it bulk} mixed solvent in $\overline{\Omega}_{sh}\cup\Omega_s$, $\epsilon_0$ is the vacuum permittivity, $\epsilon_x$ is the dielectric constant of the bulk solvent, $l_c = \sqrt{l_Bl_D/48}$ is a density-density correlation length \cite{Lee96}, and $l_B$ and $l_D$ are the Bjerrum and Debye lengths, respectively. The dielectric operator $\epsilon_s(l_c^2 \nabla^2-1)$ yields the permittivity of the {\it electrolyte solution} and the polarization of the {\it mixed solvent} as {\it functions} of \textbf{r} \cite{Liu20}.

\begin{table*}[t]
\centering
\begin{tabular}{cccc}
\multicolumn{4}{c}{Table 1. Values of model notations.} \\ \hline
Symbol & Meaning & \ Value & \ Unit \\ \hline
\multicolumn{1}{l}{$k_{B}$} & \multicolumn{1}{l}{Boltzmann constant} &
\multicolumn{1}{l}{$1.380649\times 10^{-23}$} & J/K \\
\multicolumn{1}{l}{$T$} & \multicolumn{1}{l}{temperature} &
\multicolumn{1}{l}{298.15} & K \\
\multicolumn{1}{l}{$e$} &
\multicolumn{1}{l}{proton charge} &
\multicolumn{1}{l}{$1.6022\times 10^{-19}$} & C \\
\multicolumn{1}{l}{$\epsilon_{0}$} & \multicolumn{1}{l}{permittivity of
vacuum} &
\multicolumn{1}{l}{$8.854187\times 10^{-14}$} & F/cm \\
\multicolumn{1}{l}{$\epsilon_{\text{H$_2$O}}$, $\epsilon_{\text{MeOH}}$} &
\multicolumn{1}{l}{dielectric constants} & \multicolumn{1}{l}{78.45, 31.93} &  \\
\multicolumn{1}{l}{$x$} &
\multicolumn{1}{l}{mixing percentage} &
\multicolumn{1}{l}{in [0, 1]} &  \\
\multicolumn{1}{l}{$\rho^0_{\text{H$_2$O}}$, $\rho^0_{0.5}$, $\rho^0_{\text{MeOH}}$} &
\multicolumn{1}{l}{pure solvent densities} & \multicolumn{1}{l}{0.9971, 0.9128, 0.7866 \cite{Her03}} & g/cm$^3$ \\
\multicolumn{1}{l}{$D_\text{NaF}$, $D_\text{NaCl}$, $D_\text{NaBr}$} &
\multicolumn{1}{l}{density gradients } &
\multicolumn{1}{l}{41.38, 46.62, 77.13 \cite{Bar85,Rei15}} &  g$^2$/(cm$^3$mol)\\
\multicolumn{1}{l}{$M_\text{NaF}$, $M_\text{NaCl}$, $M_\text{NaBr}$} &
\multicolumn{1}{l}{molar masses } &
\multicolumn{1}{l}{41.99, 58.44, 102.894} &  g/mol \\
\multicolumn{1}{l}{$r_{\text{Na}^{+}}$, $r_{\text{F}^{-}}$, $r_{\text{Cl}^{-}}$, $r_{\text{Br}^{-}}$, $r_{\text{H}_2\text{O}}$, $r_{\text{MeOH}}$} &
\multicolumn{1}{l}{radii} & \multicolumn{1}{l}{0.95, 1.36,
1.81, 1.95, 1.4, 1.915} & \AA  \\
\multicolumn{1}{l}{$O_x^\pm$} & \multicolumn{1}{l}{occupation number in $\Omega_{sh}$}
& \multicolumn{1}{l}{18 \cite{Mah11,Rud13}} & \\
\multicolumn{1}{l}{$R_{\text{Na}^+}^{0}$, $R_{\text{F}^-}^{0}$, $R_{\text{Cl}^-}^{0}$, $R_{\text{Br}^-}^{0}$} & \multicolumn{1}{l}{Born
radii in H$_2$O}  &
\multicolumn{1}{l}{1.587, 1.569, 2.199, 2.398}
& \AA  \\
\multicolumn{1}{l}{} & \multicolumn{1}{l}{in MeOH}  &
\multicolumn{1}{l}{1.783, 1.5, 2.02, 2.181}
& \AA  \\ \hline
\end{tabular}%
\end{table*}

The same derivation steps developed in \cite{Li20} apply to Eqs. (4) and (5) for an approximate and analytical solution, i.e., (i) linearization of the fourth-order PF (same as  Poisson-Bikerman used in \cite{Li20}) Eq. (4) for a general binary $(K=2)$ electrolyte $C_{z_2}A_{z_1}$ in the mixed solvent with the valences of the cation $C^{z_1+}$ and anion $A^{z_2-}$ being $z_1$ and $z_2$, respectively, (ii) determination of global solutions of the linear PF and Laplace equations in the spherical domain in Fig. 1, and (iii) determination of a unique solution of these two equations with the same set of the interface and boundary conditions proposed in \cite{Li20}. The analytical solution of  Eqs. (4) and (5) is
\begin{equation}
\phi(r) =
\begin{cases}
\frac{q_i}{4\pi\epsilon_sR_{i,x}^{Born}}+\frac{q_i}{4\pi\epsilon_sR_{i,x}^{sh}}\left(\Theta-1\right)\  &\mbox{in}\  \Omega_i,\\
\frac{q_i}{4\pi\epsilon_sr}+\frac{q_i}{4\pi\epsilon_sR_{i,x}^{sh}}\left(\Theta-1\right)\  &\mbox{in}\  \Omega_{sh},\\
\frac{q_i}{4\pi\epsilon_sr}\left[ \frac{\lambda_1^2e^{-\sqrt{\lambda_2}(r-R_{i,x}^{sh})}-\lambda_2^2e^{-\sqrt{\lambda_1}(r-R_{i,x}^{sh})}}{\lambda_1^2(\sqrt{\lambda_2}R_{i,x}^{sh}+1)-\lambda_2^2(\sqrt{\lambda_1}R_{i,x}^{sh}+1)}\right]\  &\mbox{in}\  \Omega_s,
\end{cases}
\label{formula:potential function}
\end{equation}
where $r = \vert{\bf r}\vert$,
\begin{equation}
\Theta = \frac{\lambda_1^2-\lambda_2^2}{\lambda_1^2(\sqrt{\lambda_2}R_{i,x}^{sh}+1)-\lambda_2^2(\sqrt{\lambda_1}R_{i,x}^{sh}+1)},
\end{equation}
\begin{equation}
\lambda_1 = \frac{ 1-\sqrt{1-(l_c^x)^2/(l_D^x)^2}}{2(l_c^x)^2}, \ \lambda_2 = \frac{1+\sqrt{1-(l_c^x)^2/(l_D^x)^2}}{2(l_c^x)^2},
\end{equation}
\begin{equation}
l_D^x = \left( \frac{\epsilon_s k_B T}{C_1^B((1-\Lambda_x)q_1^2-q_1 q_2)} \right)^{1/2},
\end{equation}
\begin{equation}
\Lambda_x = \frac{C_1^B(v_1-v_2)^2}{v_0\Gamma^B+v_1^2C_1^B+v_2^2C_2^B+v_3^2 C_3^B+v_4^2 C_4^B},
\end{equation}
$C_3^B =(1-x)\bar{C}_3^B$ with any mixing $x$ in [0, 1], $C_4^B = x\bar{C}_4^B$, $\bar{C}_3^B$ and $\bar{C}_4^B$ denote respectively the maximal bulk concentrations of water and methanol considered in this work, and the Debye $l_D^x$ and correlation $l_c^x$ lengths have been generalized to include all particle volumes as shown in $\Lambda_x$. All formulas are in the same form as those in \cite{Li20} generalized to include mixtures.

Since the solvation free energy of an ion $i$ varies with salt  concentrations, the Born energy
\begin{equation}
\begin{array}{l}
\frac{-q_i^2}{8\pi \epsilon_0 R_{i,x}^0}\left( 1-\frac{1}{ \epsilon_x} \right), \ R_{i,x}^0 = (1-x) R_{i,\text{H$_2$O}}^0 + x R_{i,\text{MeOH}}^0, \\
\epsilon_x = (1-x) \epsilon_{\text{H$_2$O}} + x \epsilon_{\text{MeOH}},
\end{array}
\end{equation}
in a pure mixed solvent ($C_j^B = 0$) should be modified to vary with $C_j^B\geq 0$ for $j=1,2$ \cite{Li20}. Here, the constant Born radii
\begin{equation}
R_{i,\text{H$_2$O}}^0 = \frac{-q_i^2}{8\pi \epsilon_0 \Delta H_{i,\text{H$_2$O}}^0}\left( 1-\frac{1}{ \epsilon_\text{H$_2$O}} \right)
\end{equation}
and
\begin{equation}
R_{i,\text{MeOH}}^0 = \frac{-q_i^2}{8\pi \epsilon_0 \Delta H_{i,\text{MeOH}}^0}\left( 1-\frac{1}{ \epsilon_\text{MeOH}} \right)
\end{equation}
in pure water and pure methanol are obtained from experimental hydration
Helmholtz free energies $\Delta H_{i,\text{H$_2$O}}^0$ and $\Delta H_{i,\text{MeOH}}^0$ \cite{Faw04,Pli13,Val15}, respectively, as given in Table 1 with other physical values. The effective Born radius $R_{i,x}^{Born}$ in \eqref{formula:potential function} is thus a function of $C_j^B$ that can be modeled by the simple formula \cite{Liu15}
\begin{equation}
R_{i,x}^{Born}(I) = \theta(I) R_{i,x}^0, \ \theta(I) = 1+\alpha^x_1\overline{I}^{1/2}+\alpha^x_2\overline{I}+\alpha^x_3\overline{I}^{3/2}, \label{formula:born radius}
\end{equation}
where $I = \frac{1}{2}\Sigma_j C_j^Bz_j^2$ is the ionic strength of the solution, $\overline{I} = I/$M is a dimensionless ionic strength, M is molarity, and $\alpha^x_1$, $\alpha^x_2$, and $\alpha^x_3$ are parameters for modifying the Born radius $R_{i,x}^0$ to fit experimental activity coefficients that change with $I$.

From \eqref{formula:excess chemical potential} and \eqref{formula:potential function}, we thus obtain a generalized activity coefficient
\begin{equation}
\ln \gamma_{i}^x(I) = \dfrac{q_i^2}{8 \pi \epsilon_s k_B T}\left(\dfrac{1}{R_{i,x}^{Born}(I)}-\dfrac{1}{R_{i,x}^0}+\dfrac{\Theta-1}{R_{i,x}^{sh}}\right) \label{formula:activity coefficient}
\end{equation}
for each ion $i$ in the mixed electrolyte solution and the mean activity coefficient
\begin{equation}
\ln\gamma_{\pm}^x(I)
=\frac{z_{2}}{z_{1}+z_{2}}\ln\gamma_+^x+\frac{z_{1}}
{z_{1}+z_{2}}\ln\gamma_-^x,
\end{equation}
where $+$ and $-$ denote ${\text{C}^{z_{1}+}}$ and ${\text{A}^{z_{2}-}}$, respectively.

We use three parameters $\alpha_j^{\text{H$_2$O}}$ for $j=1,2,3$ to fit the experimental activities of NaF \cite{Her03}, NaCl \cite{Bas96}, and NaBr \cite{Han93}, for example, in pure-water solutions. We then use another three $\Delta \alpha_j$ in
\begin{equation}
\alpha_j^x = \alpha_j^{\text{H$_2$O}} + x \Delta \alpha_j
\end{equation}
to predict the activities of these salts in mixed-solvent solutions for any $x$ in [0, 1]. This equation is derived from $\alpha_j^x = (1-x)\alpha_j^{\text{H$_2$O}} + x \alpha_j^\text{MeOH}$. Since the steric potential takes particle volumes and voids into account, the shell volume $V_{sh}$ of the shell domain $\Omega_{sh}$ can be determined by the steric potential
\begin{equation}
S_{sh} = \frac{v_0}{v_x}\ln\frac{O_x^\pm}{V_{sh}C^B_x} = \ln\frac{V_{sh}-v_x O_x^\pm}{V_{sh}\Gamma^B},
\end{equation}
where $O_x^\pm$ is the occupation (coordination \cite{Mah11,Rud13}) number of solvent molecules in $\Omega_{sh}$, $v_x = (1-x) v_3 + x v_4$, and $C_x^B = (1-x)\bar{C}_3^B + x\bar{C}_4^B$. The shell radius $R_{i,x}^{sh}$ is thus determined by $O_x^\pm$.

The following algorithm summarizes the proposed model with more details in numerical methods and implementation, where Steps 1 - 5 are for fitting and 6 - 8 for prediction.

\begin{table*}[t]
\centering
\begin{tabular}{l}
\textit{Algorithm for generalized Debye-H{\"u}ckel model} \\
\hline\noalign{\smallskip}
\textit{Input:} Experimental data ($\gamma_{\pm}^{x,exp}$, $C_+^B$) for cation $+$ and $x \in [0, 1]$ with $C_+^B$ in molality (m).\\
\textit{Functions:} \\
Solvent() returns $\epsilon_x$ (11), $v_x$ (18), $C_x^B$ (18) given $x$. \\
Born() returns $R_{\pm,x}^0$ (11) given $x$, $\Delta H_{\pm,3}^0$ (12), $\Delta H_{\pm,4}^0$ (13) with 3: \text{H$_2$O}, 4: \text{MeOH}. \\
m2M() converts $C_+^B$ to $C_+$ (in molarity M) given $x$.\\
Newton() solves a nonlinear eq. $f(V_{sh})=0$ from (18) for $V_{sh}$ that yields $R_{\pm,x}^{sh}$ (15). \\
LSfit() returns best $\gamma_{\pm}^{0}$ (16) fitted to $\gamma_{\pm}^{0,exp}$ by least squares with best $\alpha_j^\text{H$_2$O}$ (17) in $\theta(I)$ (14) for $j = 1,2,3$. \\
Activity() returns $\gamma_{\pm}^x$ from (16) given $\theta(I)$, $R_{\pm,x}^0$, $R_{\pm,x}^{sh}$. \\

\textit{Steps:} \\
1. [$\epsilon_3$, $v_3$, $\bar{C}_3^B$] = Solvent($x$ = 0).  \\
2. [$R_{\pm,3}^0$] = Born($\epsilon_3$, $x$ = 0). \\
3. [$C_+$] = m2M($C_+^B$, $x$ = 0) with M = $\dfrac{1000\text{m} \rho^l_x}{1000+\text{m}M_l}$ \cite{Mao08}, $l=$ NaF, NaCl, or NaBr, $\rho^l_x = \rho^0_x + \dfrac{D_l\text{m}}{1000}$ \cite{Bar85,Rei15},\\
\hspace*{0.2in} $\rho^0_x=\dfrac{(x-0.5)(x-1)}{0.5} \rho^0_3 + \dfrac{x(x-1)}{-0.25} \rho^0_{0.5} + \dfrac{x(x-0.5)}{0.5} \rho^0_4$. \\
4. [$R_{\pm,3}^{sh}$] = Newton($\bar{C}_3$, $v_3$) with $f(V_{sh}) = a V_{sh}^c - V_{sh} + b$, $a =  \Gamma^B \left( \bar{C}_3/O_3^\pm \right)^{-v_0/v_3}$, $b = v_3 O_3^\pm$, $c = 1 - (v_0/v_3)$. \\
5. [$\gamma_{\pm}^{0}(I)$, $\alpha_j^\text{H$_2$O}$] = LSfit($\gamma_{\pm}^{0,exp}$, $C_+$, $R_{\pm,3}^0$) for $j = 1,2,3$. \\
\hspace*{0.1in} 5.1. Get $\theta_k$ that yields best $\gamma_{\pm}^{0}(I_k)$ to $\gamma_{\pm}^{0,exp}(I_k)$ by alternating variation of $\theta$ from 1 for $k=1,2,...,N$ as follows: \\
\hspace*{0.4in} $\theta_k = 1$, $\gamma_{\pm}^0(I_k) = 1$, $n = 1$, while $\left(\left| \gamma_{\pm}^0(I_k) - \gamma_{\pm}^{0,exp}(I_k) \right|> 0.003\right)$ do \{$\theta_k = \theta_k + (-1)^n10^{-4}n$, \\
\hspace*{0.4in} [$\gamma_{\pm}^0(I_k)$] = Activity($\theta_k$, $R_{\pm,3}^0$, $R_{\pm,3}^{sh}$), $n = n+1$\}. \\
\hspace*{0.1in} 5.2. Solve $A \boldsymbol{\alpha} = \boldsymbol{b}$ from (14) for $\boldsymbol{\alpha} = \left[\begin{matrix}
		\alpha_1\\
		\alpha_2\\
		\alpha_3
		\end{matrix}\right]$ with
$A =  \left[\begin{matrix}
		\bar{I}_i^{\frac{1}{2}} \ \bar{I}_i \ \bar{I}_i^{\frac{3}{2}}\\
		\bar{I}_j^{\frac{1}{2}} \ \bar{I}_j \ \bar{I}_j^{\frac{3}{2}}\\
		\bar{I}_k^{\frac{1}{2}} \ \bar{I}_k \ \bar{I}_k^{\frac{3}{2}}
		\end{matrix}\right]$,
$\boldsymbol{b} = \left[\begin{matrix}
		\theta_i-1\\
		\theta_j-1\\
		\theta_k-1
		\end{matrix}\right]$, $i=1,\dots,N$, $j=i+1,\dots,N$, \\
\hspace*{0.4in} $k=j+1,\dots,N$. The total number of $\boldsymbol{\alpha}$s (all combinatorial $i$, $j$, and $k$) is $N_c=N(N-1)(N-2)/6$. \\
\hspace*{0.1in} 5.3. [$\gamma_{\pm, i}^0(I_k)$] = Activity($\theta_{k,i}$, $R_{\pm,3}^0$, $R_{\pm,3}^{sh}$) with $\theta_{k,i} = 1 + \boldsymbol{\alpha}_i(1)I_k^{\frac{1}{2}} + \boldsymbol{\alpha}_i(2)I_k + \boldsymbol{\alpha}_i(3)I_k^{\frac{3}{2}}$, $k=1,\dots,N$, $i=1,\dots,N_c$. \\
\hspace*{0.1in} 5.4. Error$(i) = \sum_{k=1}^N \left( \gamma_{\pm, i}^0(I_k) - \gamma_{\pm}^{0,exp}(I_k) \right)^2$, $i=1,\dots,N_c$. Set $n=i$ with Error$(i)$ being the minimum. \\
\hspace*{0.1in} 5.5. $\gamma_{\pm}^{0}(I) = \gamma_{\pm, n}^0(I)$, $\alpha_j^\text{H$_2$O} = \boldsymbol{\alpha}_n(j)$. \\
6. [$\epsilon_x$, $v_x$, $C_x^B$] = Solvent($x \ne 0$). \\
7. [$R_{\pm,x}^0$] = Born($\epsilon_x$, $x$), [$C_x$] = m2M($C_+^B$, $x$), [$R_{\pm,x}^{sh}$] = Newton($C_x$, $v_x$). \\
8. [$\gamma_{\pm, i}^x(I)$] = Activity($\theta(I)$, $R_{\pm,x}^0$, $R_{\pm,x}^{sh}$) with $\theta(I)$ in (14), $\alpha_j^x$ in (17), $\Delta \alpha_j$ being guessed. \\
\textit{Output:} [$\gamma_{\pm, i}^x(I)$, $\alpha_j^{\text{H$_2$O}}$, $\Delta \alpha_j$, $\theta(I)$] with $x \in [0, 1]$. \\
\noalign{\smallskip}\hline
\end{tabular}
\end{table*}

%
%
\section{Results and discussion}

\begin{figure*}[t]
\centerline{\includegraphics[scale=0.35]{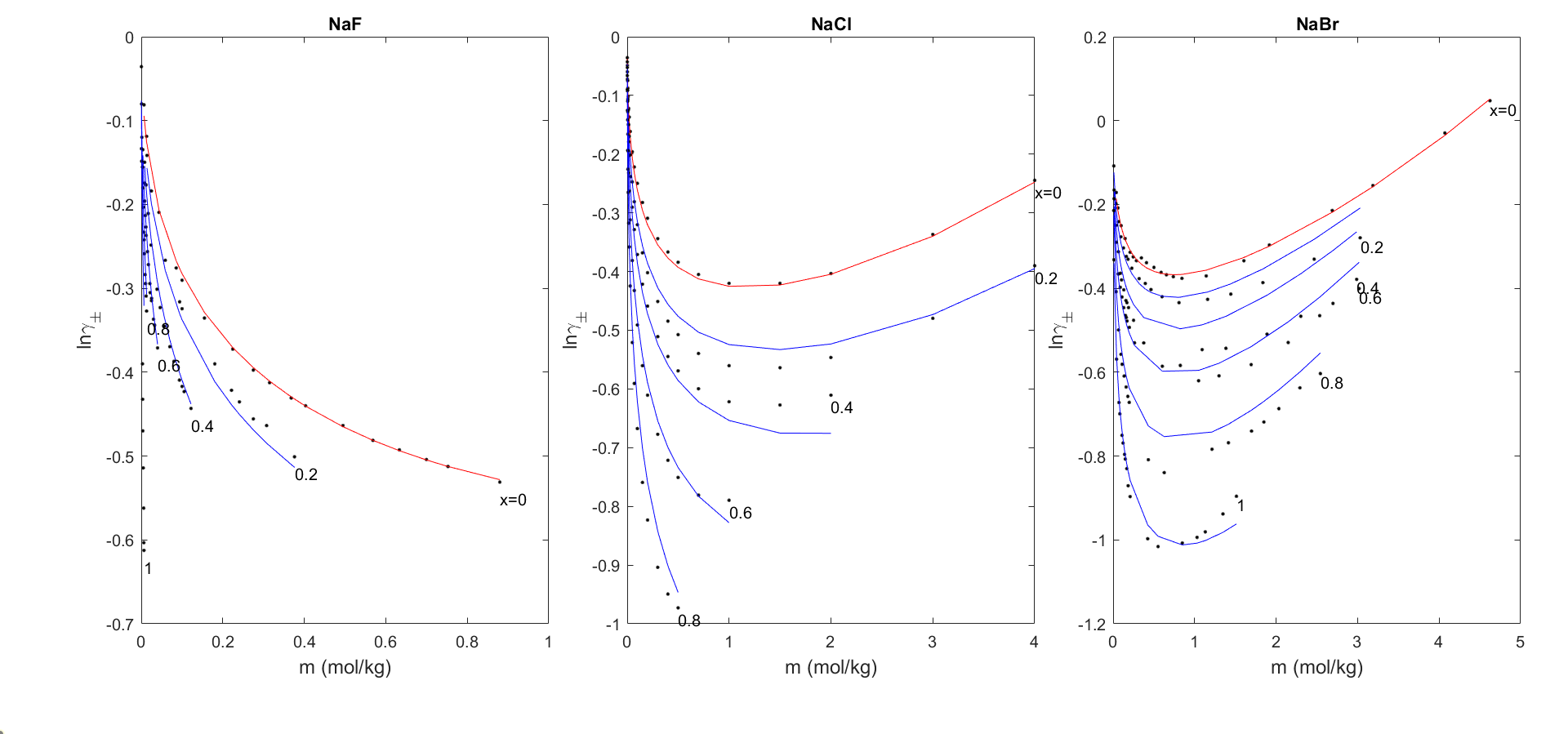}}
\caption{Mean activity coefficients of NaF, NaCl, NaBr fitted (red curves) and predicted (blue curves) by (16) with $\alpha_j^{\text{H$_2$O}}$ and $\Delta \alpha_j$ in (17) to experimental data (symbols) from \cite{Her03,Bas96,Han93} at $x=0$ (in pure water) and $x=0.2, 0.4, 0.6, 0.8, 1$ (in mixture or pure methanol), respectively.}
\end{figure*}

Figure 2 shows the mean activity coefficients of NaF, NaCl, NaBr fitted (red curves) and predicted (blue curves) by (16) with $\alpha_j^{\text{H$_2$O}}$ and $\Delta \alpha_j$ in (17) to experimental data (symbols) from \cite{Her03,Bas96,Han93} at $x=0$ (in pure water) and $x=0.2, 0.4, 0.6, 0.8, 1$ (in mixture or pure methanol), respectively. The values of $\alpha_j^{\text{H$_2$O}}$ and $\Delta \alpha_j$ for $j=1, 2, 3$ are given in Table 2 and show the significant order $\left| \alpha_1^{\text{H$_2$O}} \right| > \left| \alpha_2^{\text{H$_2$O}} \right| > \left| \alpha_3^{\text{H$_2$O}} \right|$ and $\left| \Delta \alpha_1 \right| > \left| \Delta \alpha_2 \right| > \left| \Delta \alpha_3 \right|$, which implies the order $\left| \alpha_1^x \right| > \left| \alpha_2^x \right| > \left| \alpha_3^x \right|$ as well from (17). This numerical order gives mathematical hints to these parameters for further use of our model in different conditions or for other electrolyte systems.

\begin{table*}[t]
\centering
\begin{tabular}{c|ccc|ccc|ccc}
\multicolumn{10}{c}{Table 2. Values of $\alpha_j^{\text{H$_2$O}}$ and $\Delta \alpha_j$ in (17) for NaF, NaCl, and NaBr activities in Fig. 2.} \\ \hline
\multicolumn{1}{c}{} &  \multicolumn{3}{c}{NaF} & \multicolumn{3}{c}{NaCl} & \multicolumn{3}{c}{NaBr} \\
\hline
$j$ & 1 & 2 & 3 & 1 & 2 & 3 & 1 & 2 & 3 \\
$\alpha_j^{\text{H$_2$O}}$ & 0.0224 & 0.0099 & -0.0050 & 0.0224 & -0.0113 & -0.0005 & 0.0242 & -0.0223 & 0.0009 \\
$\Delta \alpha_j$ & 0.06 & -0.01 & 0.005 &
0.068 & -0.0017 & -0.0002 & 0.027 & -0.004 & -0.0005\\
\hline
\end{tabular}
\end{table*}

Figure 3 shows that the factor $\theta$ in the effective Born
radius $\theta R_{i,x}^{0}$ of ion $i$ in (14) varies non-monotonically with the concentration of NaF, NaCl, and NaBr with different curvatures due to different sizes of anions in these salts.
It also varies with the percentage $x$ (= 0 for red solid curves and 0.2, 0.4, 0.6, 0.8, 1 for others) of methanol in mixtures. Therefore, the parameters $\alpha_j^x$ have physical meaning in Born energy and their values from Table 2 are in accord with experimental solvation energies of these ions in water-methanol mixtures \cite{Mar83,Hef88}, i.e., the Born energy of these anions in H$_2$O-MeOH mixture is larger than in pure H$_2$O in the same conditions, see e.g. Fig. 3 in \cite{Hef88}.

\begin{figure*}[t]
\centerline{\includegraphics[scale=0.35]{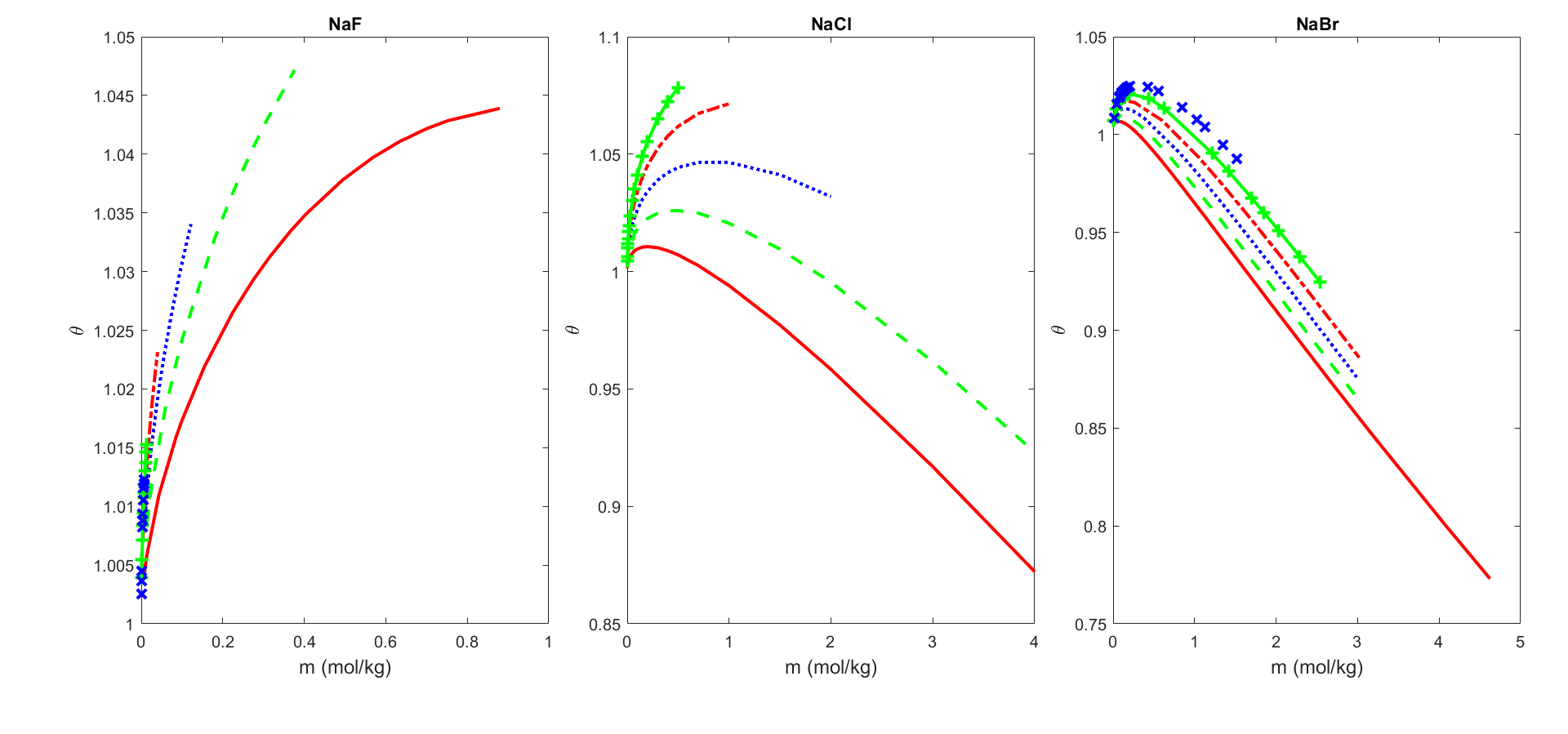}}
\caption{Variation of $\theta$ in the effective Born
radius $\theta R_{\pm,x}^{0}$ of ion $+$ or $-$ in NaF, NaCl, NaBr with concentration and $x$ (= 0 for red solid curves and 0.2, 0.4, 0.6, 0.8, 1 for others). These curves correspond to those in Fig. 2.}
\end{figure*}

We make some more remarks on the results, model, and algorithm as follows:

(i) Our model can fit any set of experimental activity data points as shown by the red curves in Fig. 2 with only 3 parameters ($\alpha_j^{\text{H$_2$O}}$, $j=1,2,3$) for which their values are determined automatically (not manually) by the algorithm. It nevertheless requires many precise physical values from experimental sources as shown in Table 1 and the algorithm.

(ii) We manually adjusted the values of $\Delta \alpha_j$ in Table 2 for prediction. In addition to the properties of significant order and Born energy, these values are not arbitrary but can be verified with experimental data. For example, $\alpha_1^\text{MeOH} = 0.0242 + 0.027  = 0.0512 \approx 0.0501$ for NaBr by (17) at $x=1$ (in pure MeOH), where $0.0242 = \alpha_1^{\text{H$_2$O}}$ (in pure H$_2$O) and $0.027 = \Delta \alpha_1$ are from Table 2, and 0.0501 is from fitting MeOH data (not shown). This implies that the value 0.027 is reasonable and verifiable.

(iii) As shown and discussed in \cite{Li20,Liu15,Liu18,Liu19}, the principal determinant of ionic activities by (15) is the effective Born radius $R_{i,x}^{Born}(I)$ (or the ion domain $\Omega_i$ in Fig. 1) due to the singular charge $q_{i}\delta(\mathbf{r}-\mathbf{0})$ of the
ion, which is infinite at $\mathbf{0}$ and thus critically affects $\gamma_\pm^x$. The secondary part in (15) is $R_{i,x}^{sh}$ that defines the shell volume of $\Omega_{sh}$ (18) in which the electric potential function $\phi(r)$ in (6) decreases exponentially to the solvent domain $\Omega_s$, see e.g. Fig. 6 in \cite{Li20}. The tertiary part is $\phi(r)$ in $\Omega_s$ that is derived by the PF Eq. (4). Therefore, our model describes the effects of ion and solvent sizes, ion-ion and ion-solvent interactions, and solution permittivity in this order of 3 determinants and of 3 subdomains.

(iv) We chose and fixed the coordination number $O_x^\pm = 18$ in (18) for simplicity to produce all the above results because of its secondary effect on $\gamma_\pm^x$. Its value can be chosen more precisely from experimental sources for different ions in different conditions \cite{Mah11,Rud13}. It can also be changed to a more specific form $O_x^\pm = (1-x)O_\text{H$_2$O}^\pm + x O_\text{MeOH}^\pm$ with changed $v_x$ and experimental $O_\text{H$_2$O}^\pm$ and $O_\text{MeOH}^\pm$, which makes (18) and Newton() in the algorithm more complicated for implementation.

(v) The activity Eq. (16) is derived from the first principle volume Eq. (1) which is a foundational proposition of our theory that defines the steric potential in (2) in terms of voids. The steric potential is thus a mean-field summary of all kinds of interactions between any pair of particles in a system such as Coulomb (long range), van der Waals (short), or Lennard-Jones (short) interactions \cite{Liu20} that produce the voids and hence the pressure of the system. Therefore, Eq. (16) does not need any mixing and combining rules (yielding more empirical parameters \cite{Kon20}) for these short-range interactions, and can apply to systems under variable temperature or pressure condition \cite{Li20}. Furthermore, Eq. (16) accounts for variable permittivity of electrolyte solutions with a dielectric function $\epsilon(r) = 1$ in $\Omega_i$ \cite{Li20}, $= \epsilon_x$ in $\Omega_{sh}$, and $= \left( \epsilon_\text{H$_2$O}C_3(r) + \epsilon_\text{MeOH}C_4(r) \right)/ \left( C_3^B + C_4^B \right)$ in $\Omega_s$ \cite{Li20}.

(vi) The model is usually expressed in dimensionless form in implementation with the scaling factors $s_1 = e/(k_B T)$ and $s_2 = \text{\AA}^2 e^2/(k_B T)$ for the potential $\phi(r)$ and concentration $C_i(r)$ variables, i.e., $s_1\phi$ and $s_2C_i$ are dimensionless.

Our code of the algorithm is accessible at https://github.com/JinnAIGroup for verification and further development.

%
%
\section{Conclusion}
We proposed a generalized Debye-H{\"u}ckel model for calculating and studying the activity of electrolytes in water-methanol mixtures for any number of salt types with arbitrary percentage (mole fraction) of methanol. The model is based on the Poisson-Fermi theory that accounts for the effects of (i) non-uniform sizes of ions and solvents, (ii) short and long interactions between ion and solvent or different ions or different solvents by mean-field {\it steric} and electric potentials, and (iii) non-uniform and size-dependent permittivity of the mixed solution.

We also proposed an algorithm to implement the model that can {\it automatically} and well  {\it fit} any set of experimental activity coefficients with corresponding salt concentrations using only 3 empirical parameters that show clear {\it physical meaning} in terms of Born energy and the significant order of their values for verification and {\it numerical hints} for further applications to other electrolyte systems. Based on these parameters, the algorithm can also {\it predict} the activity of mixtures using another 3 parameters for any mole fraction of a solvent to another solvent. Again, the later 3 parameters have the same physical meaning and significant order, and are verifiable with experimental data.

Our model and algorithm with the same parameters can straightforwardly apply to other electrolyte systems for both fitting and prediction under different conditions such as temperature or pressure.

\section*{Acknowledgements}
We thank Simon Müller at Hamburg University of Technology for motivating us to study mixtures and providing some experimental data. This work was supported by the Ministry of Science and Technology, Taiwan, through grant MOST
111-2115-M-007-010 (to CLL) and 109-2115-M-007-011-MY2 (JLL).

 %
 %


\begin{thebibliography}{99}

\bibitem{Stu12} Stumm, W., Morgan, J. J. (2012). Aquatic chemistry: chemical equilibria and rates in natural waters. John Wiley \& Sons.

\bibitem{Fra96} Franks, F. T., Ives, D. J. G. (1966). The structural properties of alcohol–water mixtures. Quarterly Reviews, Chemical Society, 20(1), 1-44.

\bibitem{Kir07} Kirchner, B. (2007). Theory of complicated liquids: Investigation of liquids, solvents and solvent effects with modern theoretical methods. Physics Reports, 440(1-3), 1-111.

\bibitem{Voi11} Voigt, W. (2011). Chemistry of salts in aqueous solutions: Applications, experiments, and theory. Pure and Applied Chemistry, 83(5), 1015-1030.

\bibitem{Row15} Rowland, D., Königsberger, E., Hefter, G., May, P. M. (2015). Aqueous electrolyte solution modelling: Some limitations of the Pitzer equations. Applied Geochemistry, 55, 170-183.

\bibitem{Ver16} Vera, J. H., Wilczek-Vera, G. (2016). Classical Thermodynamics of Fluid Systems: Principles and Applications. Crc Press.

\bibitem{Wil17} Wilhelmsen, Ø., et al. (2017). Thermodynamic modeling with equations of state: present challenges with established methods. Industrial \& Engineering Chemistry Research, 56(13), 3503-3515.

\bibitem{Kon18} Kontogeorgis, G. M., Maribo-Mogensen, B., Thomsen, K. (2018). The Debye-Hückel theory and its importance in modeling electrolyte solutions. Fluid Phase Equilibria, 462, 130-152.

\bibitem{Kon20} Kontogeorgis, G. M., Liang, X., Arya, A., Tsivintzelis, I. (2020). Equations of state in three centuries. Are we closer to arriving to a single model for all applications?. Chemical Engineering Science: X, 7, 100060.

\bibitem{Fra10} Fraenkel, D. (2010). Simplified electrostatic model for the thermodynamic excess potentials of binary strong electrolyte solutions with size-dissimilar ions. Molecular Physics, 108(11), 1435-1466.

\bibitem{Li20} Li, C. L., Liu, J. L. (2020). Generalized Debye-Hückel equation from Poisson-Bikerman theory. SIAM Journal on Applied Mathematics, 80(5), 2003-2023.

\bibitem{Liu15} Liu, J. L., Eisenberg, B. (2015). Poisson–Fermi model of single ion activities in aqueous solutions. Chemical Physics Letters, 637, 1-6.

\bibitem{Liu18} Liu, J. L., Eisenberg, B. (2018). Poisson-Fermi modeling of ion activities in aqueous single and mixed electrolyte solutions at variable temperature. The Journal of Chemical Physics, 148(5), 054501.

\bibitem{Liu19} Liu, J. L., Li, C. L. (2019). A generalized Debye-Hückel theory of electrolyte solutions. AIP Advances, 9(1), 015214.

\bibitem{Liu13} Liu, J. L. (2013). Numerical methods for the Poisson–Fermi equation in electrolytes. Journal of Computational Physics, 247, 88-99.

\bibitem{Liu13a} Liu, J. L., Eisenberg, B. (2013). Correlated ions in a calcium channel model: a Poisson–Fermi theory. The Journal of Physical Chemistry B, 117(40), 12051-12058.

\bibitem{Liu14} Liu, J. L., Eisenberg, B. (2014). Poisson-Nernst-Planck-Fermi theory for modeling biological ion channels. The Journal of Chemical Physics, 141(22), 22D532.

\bibitem{Liu20} Liu, J. L., B. Eisenberg (2020). Molecular mean-field theory of ionic solutions: a Poisson-Nernst-Planck-Bikerman model. Entropy, 22, 550.

\bibitem{Huc25} Hückel, E. (1925). Zur theorie konzentrierterer wässeriger Lösungen starker elektrolyte. Phys. Z, 26, 93-147.

\bibitem{Her03} Hernández-Luis, F., Vázquez, M. V., Esteso, M. A. (2003). Activity coefficients for NaF in methanol-water and ethanol-water mixtures at 25 C. Journal of Molecular Liquids, 108(1-3), 283-301.

\bibitem{Bas96} Basili, A., Mussini, P. R., Mussini, T., Rondinini, S. (1996). Thermodynamics of the cell: $\{\text{Na}_x\text{Hg}_{1-x}|\text{Bas96} (m)|\text{AgCl}| Ag\}$ in (methanol+ water) solvent mixtures. The Journal of Chemical Thermodynamics, 28(8), 923-933.

\bibitem{Han93} Han, S., Pan, H. (1993). Thermodynamics of the sodium bromide-methanol-water and sodium bromide-ethanol-water two ternary systems by the measurements of electromotive force at 298.15 K. Fluid Phase Equilibria, 83, 261-270.

\bibitem{Faw04} Fawcett, W. R. (2004). Liquids, solutions, and interfaces: From classical macroscopic descriptions to modern microscopic details. Oxford University Press.

\bibitem{Pli13} Pliego Jr, J. R., Miguel, E. L. (2013). Absolute single-ion solvation free energy scale in methanol determined by the lithium cluster-continuum approach. The Journal of Physical Chemistry B, 117(17), 5129-5135.

\bibitem{Val15} Valiskó, M., Boda, D. (2015). Unraveling the behavior of the individual ionic activity coefficients on the basis of the balance of ion–ion and ion–water interactions. The Journal of Physical Chemistry B, 119(4), 1546-1557.

\bibitem{Lee96} Lee, B. P., Fisher, M. E. (1996). Density fluctuations in an electrolyte from generalized Debye-Hueckel theory. Physical Review Letters, 76(16), 2906.

\bibitem{Bar85} Barthel, J., Neueder, R., Lauermann, G. (1985). Vapor pressures of non-aqueous electrolyte solutions. Part 1. Alkali metal salts in methanol. Journal of Solution Chemistry, 14(9), 621-633.

\bibitem{Rei15} Reiser, S., Horsch, M., Hasse, H. (2015). Density of methanolic alkali halide salt solutions by experiment and molecular simulation. Journal of Chemical \& Engineering Data, 60(6), 1614-1628.

\bibitem{Mah11} Mähler, J., Persson, I. (2012). A study of the hydration of the alkali metal ions in aqueous solution. Inorganic chemistry, 51(1), 425-438.

\bibitem{Rud13} Rudolph, W. W., Irmer, G. (2013). Hydration of the calcium (II) ion in an aqueous solution of common anions (ClO 4-, Cl-, Br-, and NO 3-). Dalton Transactions, 42(11), 3919-3935.

\bibitem{Mao08} Mao, S., Duan, Z. (2008). The P, V, T, x properties of binary aqueous chloride solutions up to T= 573 K and 100 MPa. The Journal of Chemical Thermodynamics, 40(7), 1046-1063.

\bibitem{Mar83} Marcus, Y. (1983). Thermodynamic functions of transfer of single ions from water to nonaqueous and mixed solvents: Part I-Gibbs free energies of transfer to nonaqueous solvents. Pure and Applied Chemistry, 55(6), 977-1021.

\bibitem{Hef88} Hefter, G. T., McLay, P. J. (1988). The solvation of fluoride ions. I. Free energies for transfer from water to aqueous alcohol and acetonitrile mixtures. Journal of Solution Chemistry, 17(6), 535-546.

\end{thebibliography}
\end{document}